# Atmospheric Profiles at the Southern Pierre Auger Observatory and their Relevance to Air Shower Measurement

J. Blümer, R. Engel, D. Gora, P. Homola, B. Keilhauer, H. Klages, J. Pekala, M. Risse, M. Unger, B. Wilczynska, H. Wilczynski for the Pierre Auger Collaboration*
Presenter: B. Keilhauer (bianca.keilhauer@ik.fzk.de), ger-keilhauer-B-abs2-he14-poster

The dependence of atmospheric conditions on altitude and time have to be known at the site of an air shower experiment for accurate reconstruction of extensive air showers and their simulations. The height-profile of atmospheric depth is of particular interest as it enters directly into the reconstruction of longitudinal shower development and of the primary energy and mass of cosmic rays. For the southern part of the Auger Observatory, the atmosphere has been investigated in a number of campaigns with meteorological radio soundings and with continuous measurements of ground-based weather stations. Focussing on atmospheric depth and temperature profiles, temporal variations are described and monthly profiles are developed. Uncertainties of the monthly atmospheres that are currently applied in the Auger reconstruction are discussed.

## 1. Introduction

The Pierre Auger Observatory measures extensive air showers (EAS) induced by ultra-high energy cosmic rays using a hybrid technique [1]. One detection method is the registration of secondary particles of EAS at ground with water Cherenkov tanks [2]. The second technique is the observation of the longitudinal development of EAS with fluorescence telescopes [3]. Especially for the detection and reconstruction of fluorescence light emission of EAS, knowledge of actual atmospheric conditions at the site of the experiment is necessary.

Therefore, several meteorological systems have been installed at the site of the Pierre Auger Observatory and additionally, data from publicly available databases of atmospheric measurements are used. The data recorded at the site are compared to the US Standard Atmosphere 1976 (US-StdA), which had been commonly used.

The longitudinal development of EAS can be described by the number of particles at a given atmospheric depth. The atmospheric depth at which a shower exhibits its maximum, $X_{\max}$, is well correlated with the mass of the primary particle. However, using the fluorescence technique for detecting EAS, these quantities cannot be observed directly. The fluorescence telescopes (FD) observe the light within a fixed field of view. Thus, the simulated shower profiles have to be transformed from a description based on vertical atmospheric depth to geometrical height. For a physical interpretation of detected EAS events, the conversion has to be done vice versa. Therefore, the transformation between atmospheric depth and geometrical altitude is a crucial point in the simulation and reconstruction of EAS and the relation between atmospheric depth and height follows from the air density profile [4]. Atmospheric conditions also have a major impact on the fluorescence emission process itself, and on the details of light propagation from the emitting region to the telescope.

## 2. Atmospheric conditions at the southern Pierre Auger Observatory

Since August 2002, meteorological radio soundings have been performed in several campaigns near Malargüe, Argentina. The radiosondes are launched above the site of the experiment on helium-filled balloons. A set of data is taken about every 20 m during ascent up to 25 km a.s.l. in average. Despite changing wind conditions, the radiosondes stay mostly directly above the array up to 10 km a.s.l., covering the more interesting part of the profiles for EAS development. More than 100 atmospheric profiles, including data for temperature, pressure, relative humidity, and wind speed and direction, were collected. Day-night variations are very small in this area. Only temperature changes up to 10 K may occur in the lowest 1000 m above ground which is roughly at 1420 m a.s.l. in our case. The important profile of atmospheric depth hardly changes on day-night time scales. From day to day, the extent of variation is strongly seasonal dependent. During austral summer, the conditions are much more stable than during winter. At ground, differences in atmospheric depth up to 5 g cm$^{-2}$ have been found which are related to pressure variations. At higher altitudes, between 6 and 12 km a.s.l., even variations of 10 - 15 g cm$^{-2}$ are recorded. Over a period of days, the temperature may shift by 15 K. Seasonal effects are, of course, the largest. For individual days, the difference in atmospheric depth between summer and winter can be as large as 20 g cm$^{-2}$ at ground and reach approximately 30 g cm$^{-2}$ at altitudes between 6 and 9 km a.s.l.

*Observatorio Pierre Auger, Av. San Martin Norte 304, 5613 Malargüe, Argentina, http://www.auger.org/auger-authors.pdf

Apart from these intermittent profile measurements, ground-based weather stations record temperature, pressure, relative humidity, and wind data every 5 min. At completion of the southern Observatory stations will be located at every FD building and the central laser facility. Up to now, data from 2 stations are available. While the atmospheric profiles are mainly used for simulation and reconstruction of the longitudinal shower development, the continuous ground-based data are applied in the calculation of trigger efficiencies of the water Cherenkov tanks. An example of the pressure distribution within a single day is given in Fig. 1. In Figures 2 and 3, atmospheric depth and temperature data of the year 2004 are shown. Additionally to these local measurements, information from public databases are included in our analyses. The UK Met Office, through the British Atmospheric Data Centre [5], maintains a database of atmospheric radio soundings worldwide. The stations closest to Malargüe are Cordoba and Santa Rosa, at a distance of 500 - 650 km.

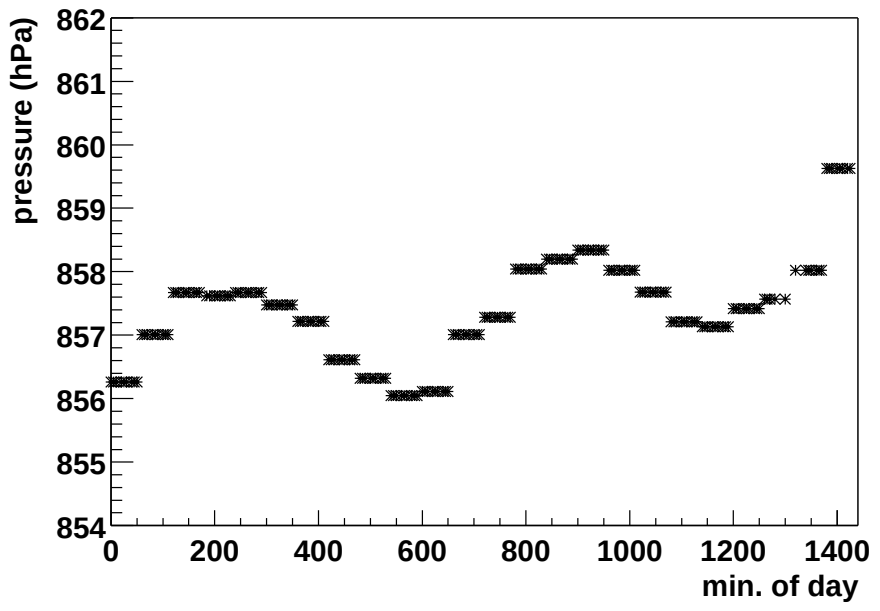

**Figure 1.** Pressure distribution at the FD building Los Leones at April 3rd, 2004.

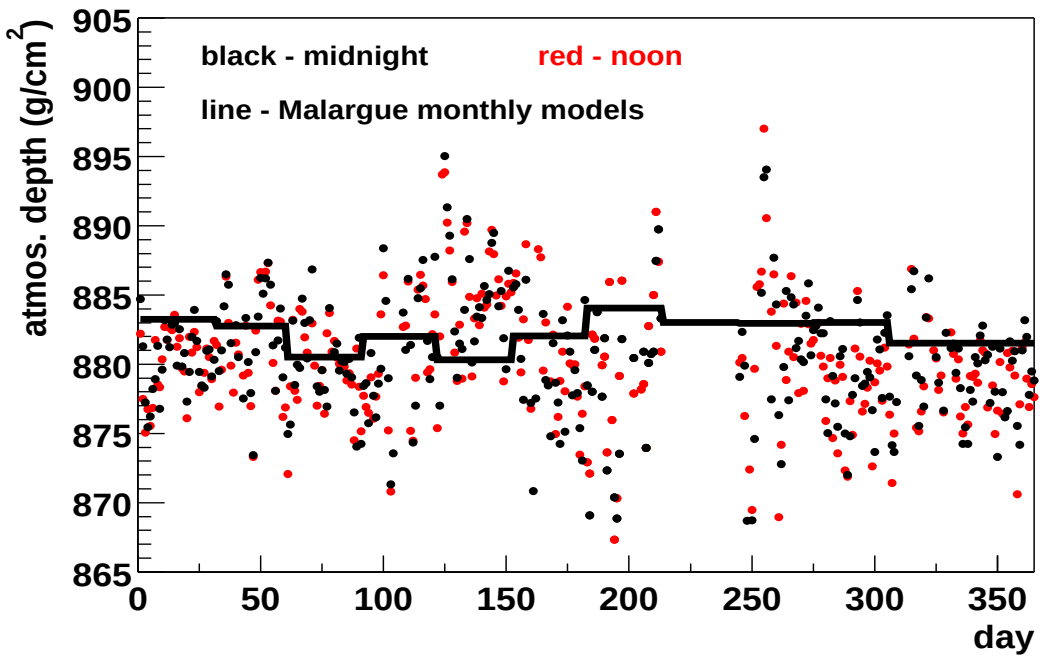

**Figure 2.** Atmospheric depth data at the FD Los Leones from the year 2004. The plotted data are recorded at noon and midnight local time. The line indicates the values of the Malargüe Monthly Models at the same altitude.

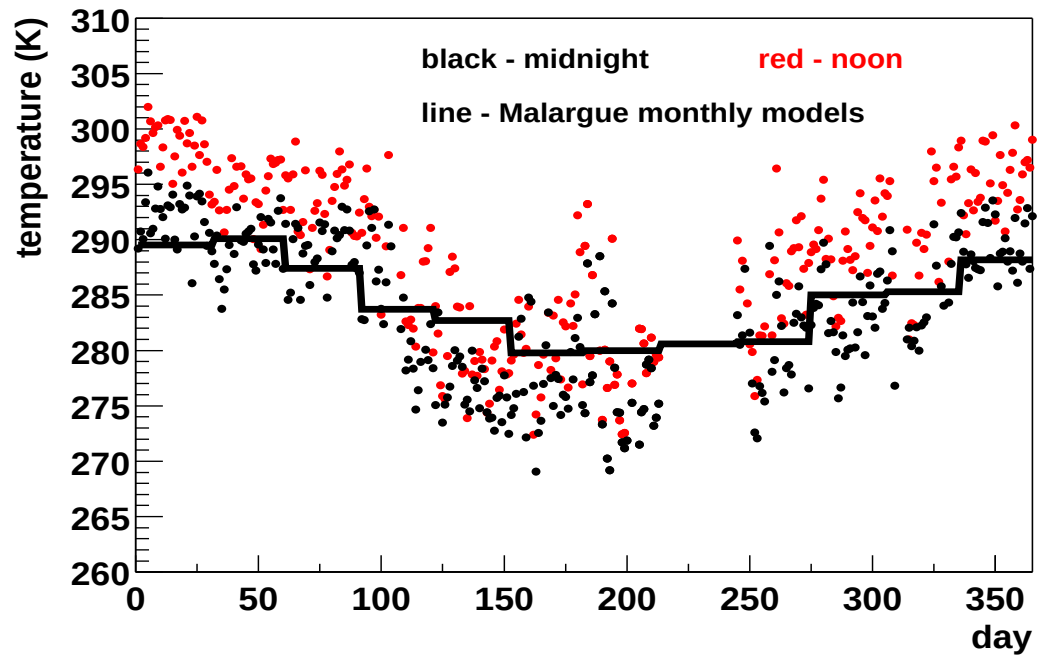

**Figure 3.** Temperature data at the FD Los Leones from the year 2004. The plotted data are recorded at noon and midnight local time. The line indicates the values of the Malargüe Monthly Models at the same altitude.

The investigation of locally measured atmospheric profiles and the comparison of them to data from the UK Met Office shows that monthly parameterisations can give a good description of the atmospheric variations. To obtain such parameterisations, once-a-day measurements of Cordoba and Santa Rosa are averaged within each month and also between these two stations. The resulting monthly profiles describe the atmosphere near

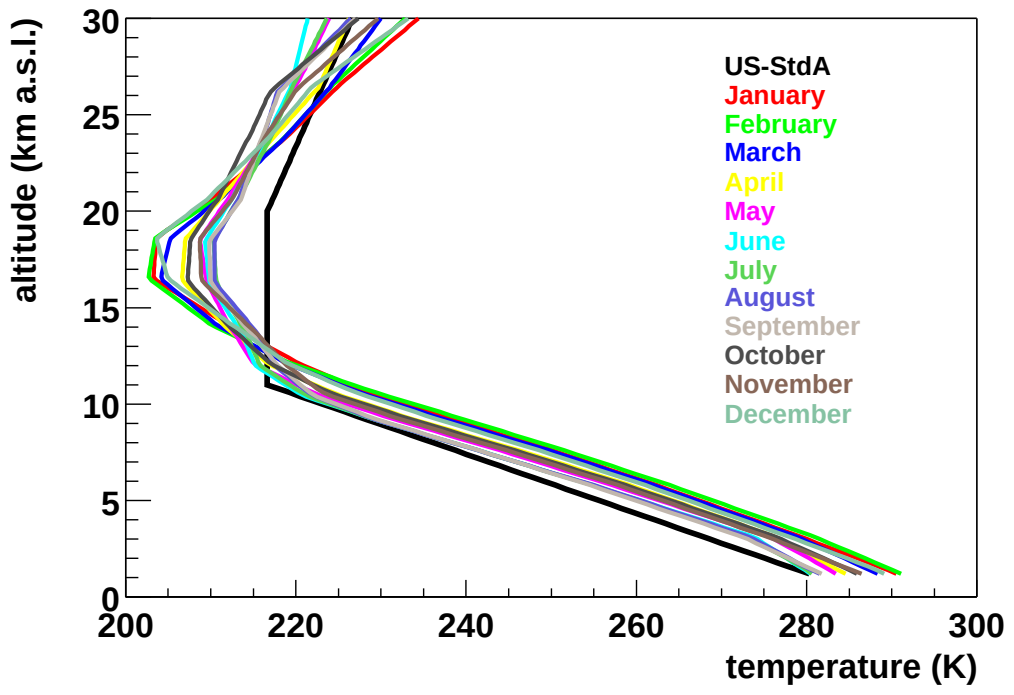

**Figure 4.** Temperature profiles of the Malargüe Monthly Models.

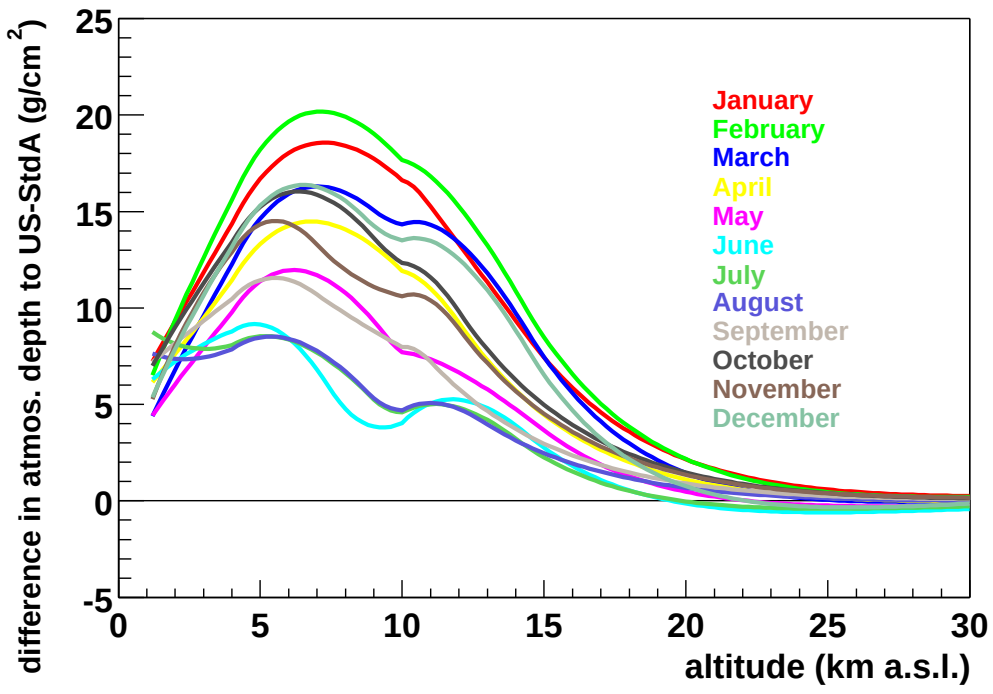

**Figure 5.** Profiles of atmospheric depth of the Malargüe Monthly Models in difference to the US-StdA.

Malargüe already quite well. Further adjustments are calculated using the information from the local radio soundings. The obtained monthly models are shown in Figures 4 and 5. These profiles, called Malargüe Monthly Models, are employed in the Auger simulation and reconstruction of EAS events [6].

## 3. Effects on longitudinal profiles

For the physical interpretation of EAS, the position of shower maximum and the energy of the primary particle are important features. Using the fluorescence technique, the shower maximum can be observed directly while the energy of the primary particle has to be deduced from the deposited ionisation energy in the atmosphere.

In Figure 6, the shower development represented by the energy deposit profile can be seen for an average of 100 Fe-ind. showers with $10^{19}$ eV and 60° incidence. The profiles are plotted versus altitude developing in the US-StdA, Malargüe February, August, and annual average atmospheres. The figure reveals that the position of shower maximum is shifted due to atmospheric conditions. To clearly demonstrate the effect, the inclination angle $\vartheta$ is chosen to be 60°, since the extent of the shift is enlarged by a factor $1/\cos\vartheta$. In particular, the Malargüe atmospheres give a systematic shift of the position of $X_{\max}$ towards higher altitudes which is equivalent to smaller values of atmospheric depth. However, not only the average position of the shower maximum is a measure of the type of the primary particle. Also the width of the distribution of this position for a large number of EAS is systematically different for proton and iron induced showers. The daily variations of the atmosphere lead to a broadening of the $X_{\rm max}$ distribution by about 25% for iron but only 4% for proton induced showers, again $\vartheta = 60°$, as compared to the expectation for the time-independent US-StdA [4].

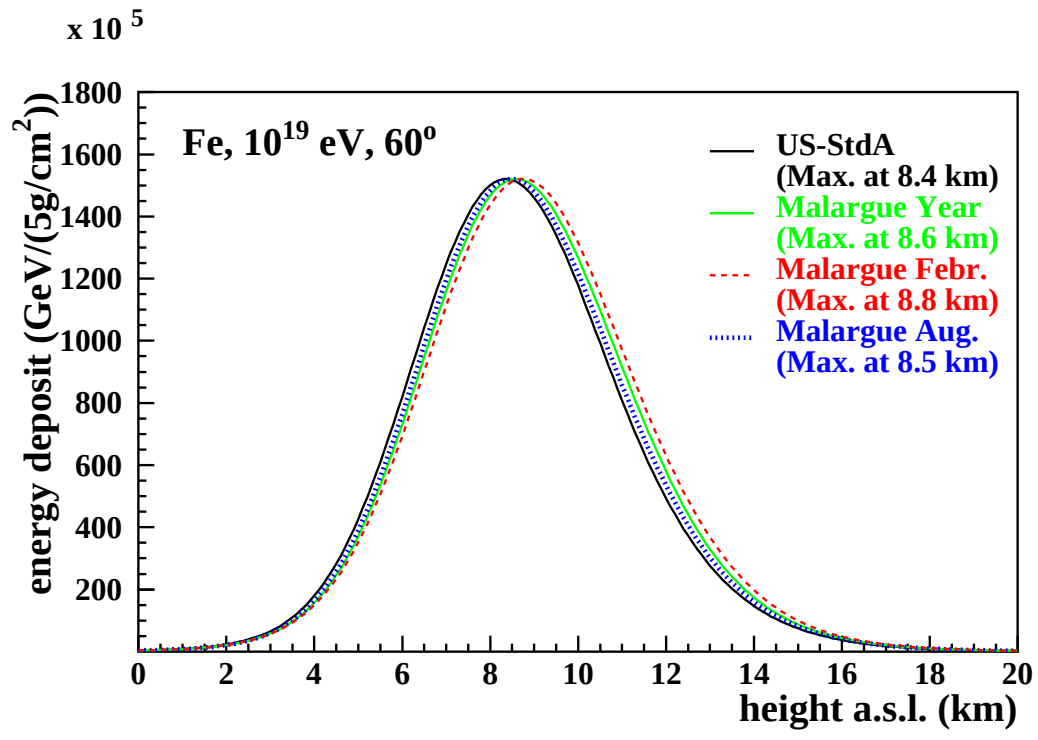

**Figure 6.** Energy deposit profile for an average of 100 EAS. The black-solid line shows the profile expected in the US-StdA. The coloured curves are for the two extrema of the Malargüe Monthly Models and the annual average.

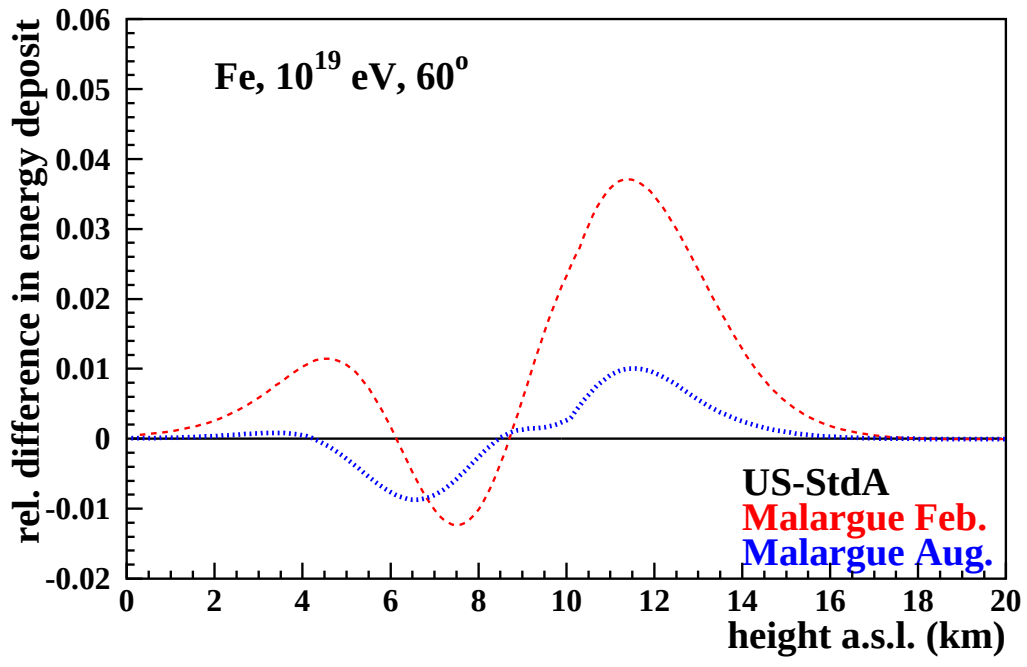

**Figure 7.** Difference of the profiles shown in Fig. 6 to the US Standard atmosphere divided by the energy deposit at shower maximum. The profiles are shifted in height to bring all shower maxima to the same position (see text).

For estimating changes in the reconstruction of the primary energy of EAS, the distortion of profiles due to atmospheric variations has to be checked. In Fig. 7, differences of two Malargüe profiles to the US-StdA are plotted. To remove obvious differences, the profiles are shifted in height such that all maxima are at the position of the maximum for the US-StdA. The variation of the total energy is negligible since the integral over each curve is small in comparison to the entire amount of deposited energy of the EAS.

## 4. Discussion of remaining uncertainties

Firstly, the influence of applying Malargüe Monthly Models instead of US-StdA is analysed with about 2700 hybrid events. The position of shower maximum is shifted on average by $\approx$ -15 g cm$^{-2}$, see Fig. 8, right panel. The change of the reconstructed primary energy is only 0.7%, see Fig. 8, left panel.

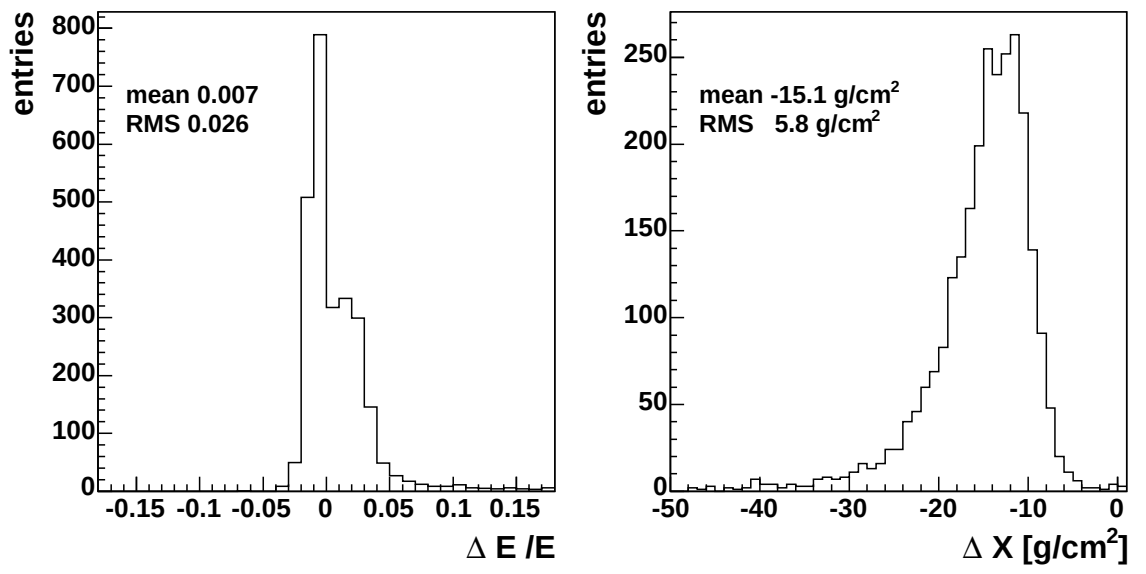

**Figure 8.** Difference of reconstruction using Malargüe Monthly Models to reconstruction in US-StdA is shown. Left: primary energy. Right: position of shower maximum.

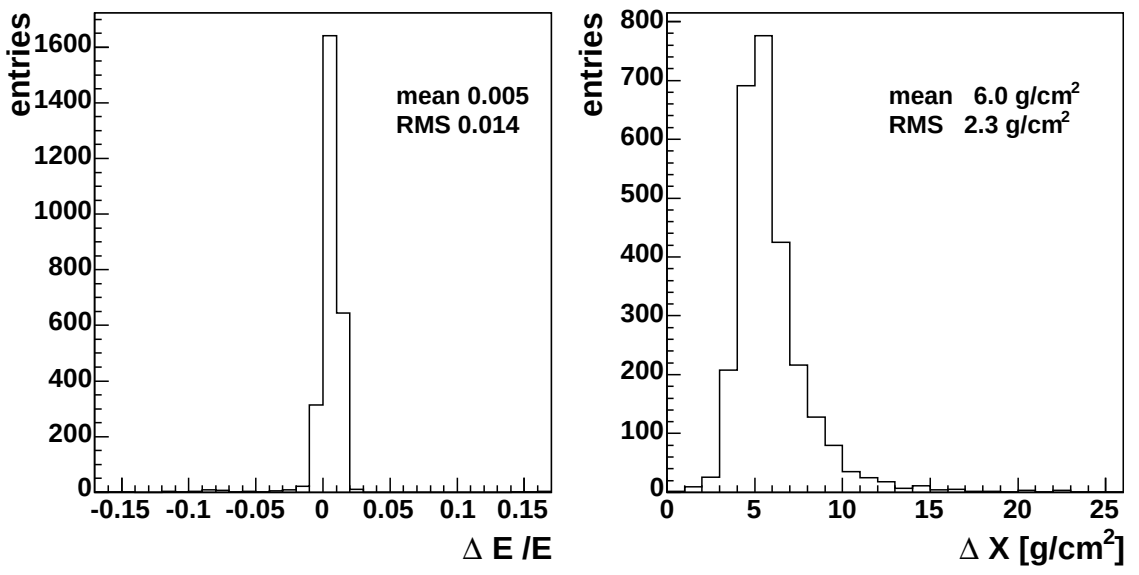

**Figure 9.** Comparison between normal reconstruction and reconstruction applying an one-$\sigma$ error to the atmospheric profiles. Left: primary energy. Right: $X_{\rm max}$ position.

Secondly, the remaining uncertainties due to day-to-day variations of the atmospheric conditions within each month have been studied using the same set of events. For this purpose, modified monthly profiles are used in the analysis that represent the uncertainty bound of one standard deviation of the Malargüe Monthly Models. The position of $X_{\rm max}$, Fig. 9, right panel, and the primary energy, Fig. 9, left panel, are compared between the normal reconstruction and the modified reconstruction. The uncertainty of the depth of maximum of EAS is about 6 g cm$^{-2}$ and of the reconstructed primary energy 0.5%. Finally it should be noted that the calculation of fluorescence photon profiles is included in the standard Auger reconstruction chain. The fluorescence photon yield is dependent on temperature and pressure according to the used atmospheric models.